\documentstyle[11pt,paspconf,epsfig]{article}

\begin{document}

\title{A Simultaneous ASCA and RXTE Long Look at the Seyfert 1 Galaxy MCG--6-30-15}

\author{J. C. Lee,  A. C. Fabian, and K. Iwasawa}
\affil{Institute of Astronomy; Madingley Road; Cambridge CB3 0HA  UK}

\author{W. N. Brandt} 
\affil{Department of Astronomy and Astrophysics; The Pennsylvania State University; 525 Davey Lab; University Park, PA 16802 USA}

\author{C. S. Reynolds }
\affil{JILA; Campus Box 440; University of Colorado; Boulder, 80309-0440 USA}




\begin{abstract}
We report on a joint RXTE and ASCA observation spanning $\sim$ 400~ks
of the  Seyfert 1 galaxy MCG--6-30-15. The low energy coverage of ASCA
coupled with the high  energy coverage of RXTE have allowed us not 
only to confirm features of Compton reflection but also to set bounds on
the abundances as a function of  reflective fraction.
\end{abstract}




\section{Introduction}

The current paradigm for AGN is a central engine consisting of an
accretion  disk surrounding a supermassive black hole (e.g. see review
by Rees 1984).  The main source of power is the release of
gravitational potential  energy as matter falls towards the central
black hole.  Much of this energy is released in the form of X-rays,
some fraction of which are reprocessed by matter in the AGN.

Careful study of X-ray reprocessing mechanisms can give much
information about the immediate environment of the accreting black
hole.  These effects of reprocessing can often be observed in the form
of emission and absorption features in the X-ray spectra of AGNs.  In
Seyfert 1 nuclei, approximately half of the X-rays are {\it reflected}
off the inner regions of the accretion disk.  The reflected spectrum
is complicated with features of photoabsorption, iron fluorescence and
Compton scattering (George \& Fabian 1991).  The strength, shape and
broadening of the features of the reflected spectrum are diagnostics
of the geometry, ionization state, and iron abundance of the accretion
disk.

MCG--6-30-15 is a Seyfert 1 galaxy that is both bright and nearby
($z=0.008$).  The spectral features above a few keV can be modeled
well by a power-law continuum plus reflection component that
encompasses the effects of reflection of this continuum by the inner
regions of the accretion disk.  The principle observables of this
reflection component are a strong iron fluorescence line at
$\sim~6.4$~keV and a Compton `hump' peaking at 20--30~keV.  The iron
line together with the reflection component are important diagnostics
for the geometry and physics of the X-ray continuum source.  The
strength of the emission line relative to the reflection hump depends
largely on the abundance of iron relative to hydrogen in the disk.
Disentangling the abundance from the absolute normalization of the
reflection component is an important step in constraining physical
models of AGN central regions.

Previous studies of GINGA spectra have suggested that the reflection
component is present in many Seyfert 1 AGNs (Nandra \& Pounds 1994).
Our recent observations using the Rossi X-ray Timing Explorer (RXTE)
have clearly confirmed the presence of reflection in MCG--6-30-15.

\section{Observations}
\begin{figure}
\begin{center} \leavevmode

\centerline{\epsfig{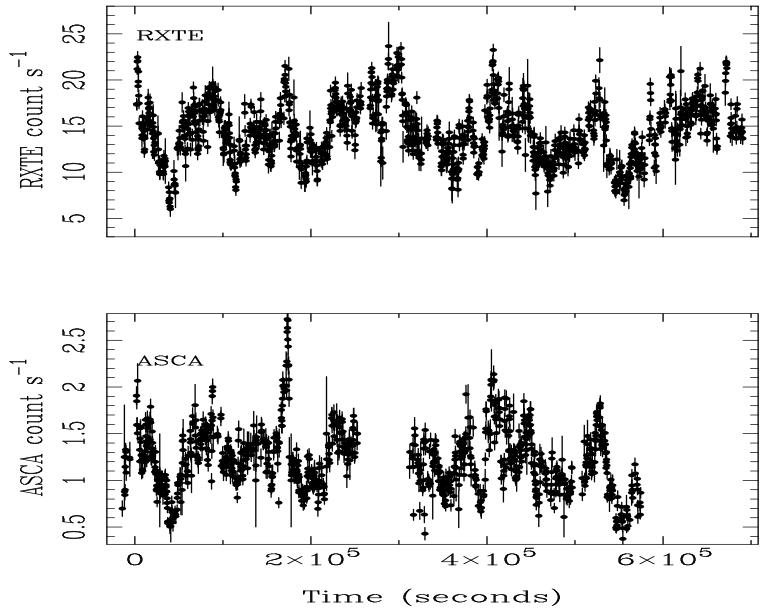}}

\end{center}

\caption[h]{MCG--6-30-15 was observed simultaneously by RXTE and ASCA
spanning $\sim$ 400~ks over the period from 1997 August 3 to 1997 August 12.
The RXTE on-source time was $\sim$ 400~ks and the ASCA on-source time was
  $\rm \sim 200~ks$.}  \label{fig-1}

\end{figure}

MCG--6-30-15 was observed by RXTE for 400~ks over the period from 1997
August 4 to 1997 August 12 by both the Proportional Counter Array
(PCA) and High-Energy X-ray Timing Experiment (HEXTE) instruments.  It
was simultaneously observed by the Advanced Satellite for
Cosmology and Astrophysics (ASCA) Solid-state Imaging Spectrometers
(SIS) for 200 ks over the period 1997 August 3 to 1997 August 10 with
a half-day gap in the middle.  We concentrate primarily on the RXTE
observation.

PCA light curves and spectra are extracted from only the top Xenon
layer using the {\sc ftools v.4.0} software. We use only combined data
from  PCUs 0, 1, and 2 since PCUs 3 and 4 are sometimes turned off due
to occasional problems with discharge.  Good time intervals were
selected to exclude any Earth or South Atlantic Anomaly (SAA) passage
occulations, and to ensure stable pointing.


 

We generate background data using {\sc pcabackest v2.0c} in order  to
estimate the internal background caused by interactions between the
radiation/particles and the detector/spacecraft at the time of
observation.  This is done by matching the conditions of observations
with those in various model files. The model files that we chose were
constructed using the VLE rate (one of the rates in PCA Standard 2
science array data that is defined to be the rate of events which
saturate the analog electronics) as the tracer of the particle
background.

The PCA response matrix for the RXTE data set was provided by the RXTE
Guest Observer Facility (GOF) at Goddard Space Flight Center.
Background models and response matrices are representative of the most
up-to-date PCA calibrations.

The net HEXTE spectra were generated by subtracting spectra from the
off-source positions from the on-source data.  Time intervals were
chosen to exclude 30 seconds prior to and following SAA  passages.
This avoids periods when the internal background is changing rapidly.
We use response matrices provided by the HEXTE team at the University
of California, San Diego.  The relative normalizations of the PCA and
the two HEXTE clusters  are allowed to vary, due to uncertainties (
$<$ about 5\% ) in the HEXTE deadtime measurement.

ASCA data reduction was carried out using {\sc ftools} version 4.0 and
4.1 with standard calibration provided by the ASCA GOF.  Detected SIS
events with a grade of 0, 2, 3 or 4 are used for the analysis.  One of
the standard data selection criteria, {\sc br earth}, elevation angle
of the source from the bright Earth rim, is found to affect little the
soft X-ray data from the SIS.  We thus use the SIS data of
approximately 231 ks from each detector for spectral analysis.  The
source counts are collected from a region centred at the X-ray peak
within $\approx$ 4 arcmin from the SIS and 5 arcmin from the GIS.  The
background data are taken from a (nearly) source-free region in the
same detector with the same observing time.

Figure~1 shows the ASCA S0 160-2700 pha-channel ($\approx$
0.6--10~keV),  and the RXTE PCA 1--129 pha channel ($\approx$
2-60~keV) background subtracted light curves.   There is a gap of
$\sim$ 60~ks in the ASCA light curve in which the satellite observed
IC4329A while MCG--6-30-15 underwent a large flare observed by RXTE.
Significant variability can be seen in both light curves on short and
long timescales.  Flare and minima events are seen to correlate
temporally in both light curves.

\begin{figure}
\begin{center} \leavevmode

\centerline{\epsfig{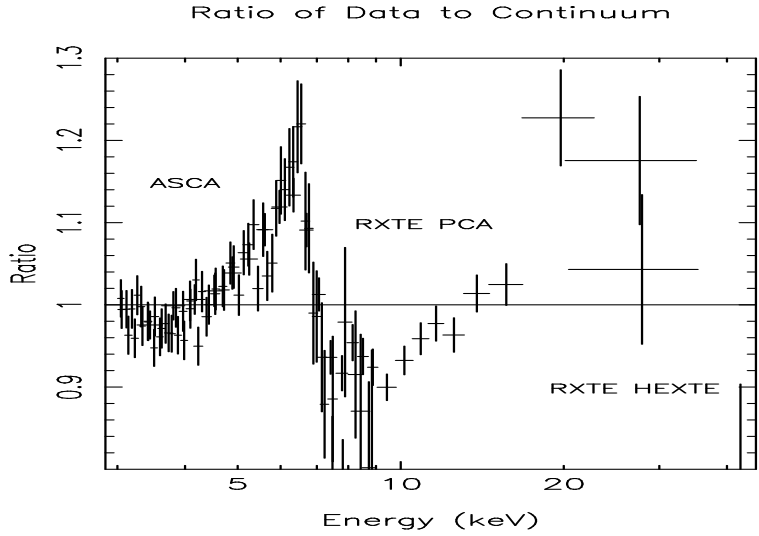}}
\caption[h]{A {\bf \it nominal} fit using a simple power-law
demonstrates the clear  existence of a redshifted broad iron line at
$\sim$ 6.0~keV and reflection component above 10~keV. } \label{fig-2}

\end{center}
\end{figure}

\begin{figure}

\begin{center} \leavevmode

\centerline{\epsfig{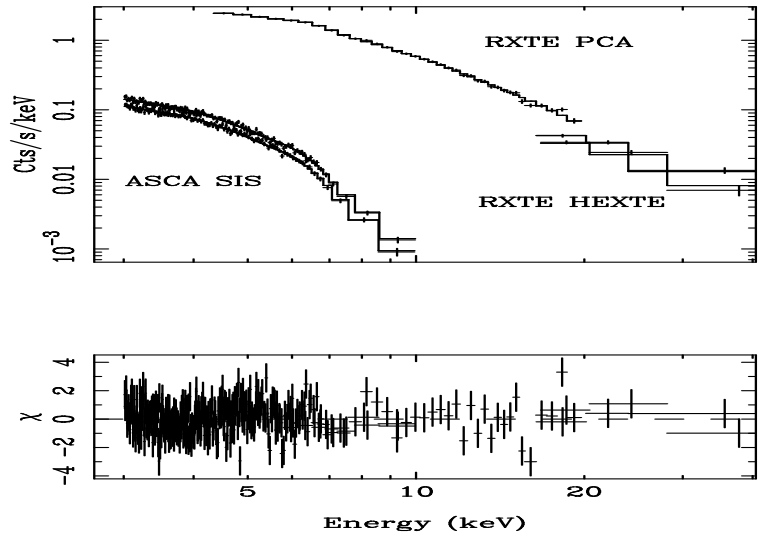}}
\caption{A multicomponent model fit that includes the reflected spectrum to all
three data sets (i.e. ASCA, RXTE PCA and HEXTE) shows that the residuals are essentially flat. } \label{fig-3}

\end{center}
\end{figure}

%

\section{Spectral Fits}

We restrict ASCA and PCA data analysis to be respectively between
3--10~keV and 3--20~keV (the PCA on-source spectrum for MCG--6-30-15 
is largely background dominated past 20~keV).  The lower energy restriction at
3~keV is selected in order that the necessity for modeling
photoelectric absorption due to Galactic ISM material, or the warm
absorber that is known to be present in this object  (i.e. Reynolds et
al. 1995) is removed.  HEXTE data is restricted to be between 16 and
50~keV in order that we may adequately model the reflection hump.  We
also include 0.5 per cent systematics to the PCA data.

\subsection{Spectral Features}
A {\it nominal} fit using a simple power law confirms the clear
existence of  a redshifted broad iron line at $\sim$ 6.0~keV and
reflection component above  10~keV as shown in Lee et al. (1998, in
press).  A plot of the ratio of  data to continuum shown in figure~2
demonstrates the significance of these  features.

\begin{figure}
\begin{center} \leavevmode

\centerline{\epsfig{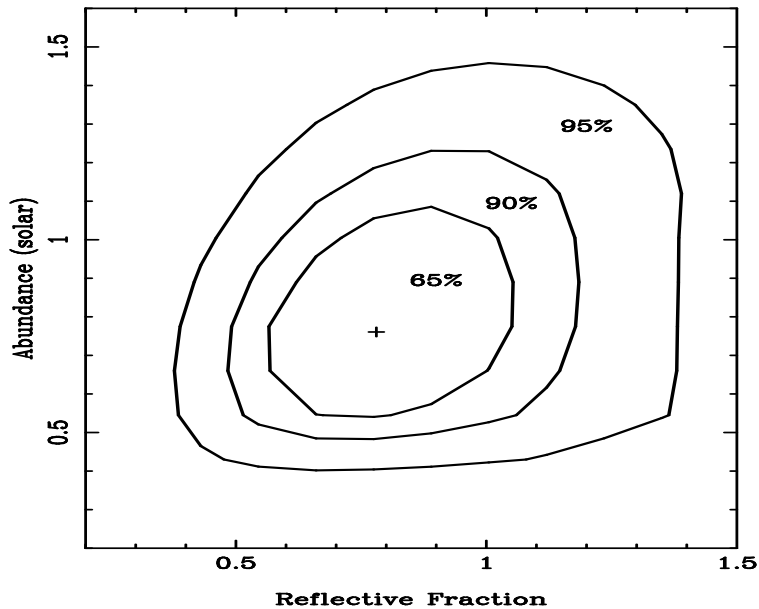}}

\caption[h]{RXTE PCA+HEXTE 65, 90, and 95 per cent confidence contours for the relationship between abundances and reflective fraction. The energy range of fit is between
3 and 50~keV. } \label{fig-4}

\end{center}
\end{figure}

As further evidence for the existence of the reflection component and
good agreement between ASCA and RXTE, figure~3 shows that the residuals
are essentially flat when all three data sets (i.e. ASCA, RXTE PCA +
HEXTE) are fit with a multicomponent model .  This model consists of a
power-law reflection component (Magdziarz \& Zdziarski 1995) to model
the primary and  reflected continuum with an additional Gaussian
component to represent the iron $\rm K \alpha$ emission.  Fitting the
RXTE data in the energy range $\rm 3-50~keV$  we obtain for a best fit
power-law slope $\Gamma$ = $2.05^{+0.13}_{-0.06}$.
The line energy is $5.94^{+0.15}_{-0.19}$~keV, with line width $\rm
\sigma$ = $0.62^{+0.20}_{-0.25}$~keV.  The equivalent width EW =
$305^{+6}_{-19}$ eV, and the reduced-$\chi^2$ for the overall fit is
0.51 for 48 degrees of freedom.

The improved S/N (as compared to GINGA) along with the broad waveband
coverage afforded by RXTE has allowed us not only to detect features
associated with Compton reflection  but also to set bounds on the
abundances as a function of reflective fraction.  Figure~4 shows that
95 per cent confidence  contours can be obtained by fitting in the
energy range 3--50~keV, with best fit values for reflective fraction
and lower  elemental abundances set equal to that of iron respectively
to be $0.78 \pm 0.31$ and  $0.76^{+0.33}_{-0.29}$ solar abundances.
The reflective fraction is defined such that its value equal to unity
implies that the X-ray source is subtending 2 $\pi$ sr of the sky
(i.e. $\frac{\Omega}{2 \pi} = 1$).

In order to test the consistency of our results, we perform fits
similar to those above on an $\sim$ 187~ks subset (from the first
portion) of the RXTE observation, and find that results are nearly
identical.  This gives us confidence that our derived parameters have
physical meaning. 
Implications of the large EW of the iron line will be address in future 
publications.

\section{Summary}
Previous studies of reflection of X--rays from optically thick cold
matter in the central region of AGNs have concluded that a reflected
spectrum exists and is observed above the primary continuum.  However,
most studies in the past have been only able to consider the iron line
alone due in part to a lack of adequate waveband coverage and / or
good spectral resolution.

The improved S/N (as compared to GINGA) along with the high energy
coverage afforded by RXTE coupled with simultaneous ASCA observations
have allowed us not only to detect such features in our observations
of MCG--6-30-15 but to set bounds on the abundances as a function of
reflective fraction (figure 4) at the 95 per cent confidence level.
This will have important consequences for  understanding the geometry,
and constraining processes in AGN central regions.

\acknowledgments

We thank all the members of the RXTE GOF for answering our inquiries
in such a timely manner, with special thanks to William Heindl and the
HEXTE team for help with HEXTE data reduction.  We also thank Keith
Jahoda for explanations of PCA calibration issues.  JCL thanks the
Isaac Newton Trust, the Overseas Research Studentship programme (ORS)
and the Cambridge Commonwealth Trust for support.  ACF thanks the
Royal Society for support. CSR thanks the National  Science Foundation
for support under grant AST9529175, and NASA for support under the
Long Term Space Astrophysics grant NASA-NAG-6337. KI and WNB thank
PPARC and NASA RXTE grant NAG5-6852 for support, respectively.

%
%

\end{document}